\documentclass[conference]{IEEEtran}
\usepackage{amsfonts,amsmath,amssymb}
\usepackage{cite}
\usepackage{url}
\usepackage{subfig}
\usepackage{multirow}
\usepackage{multicol}
\usepackage{floatrow}
\usepackage{paralist}
\usepackage{graphicx}
\usepackage{float}
\usepackage{csvsimple}
\usepackage{rotating}
\usepackage{pdflscape}
\usepackage{enumitem}
\graphicspath{{./figures/}}

\newcommand{\ie}{{i.e.}}
\newcommand{\eg}{{e.g.}}
\newcommand{\etal}{et al.}

\ifCLASSINFOpdf
\else
\fi
\begin{document}
%
\title{On Multi-Session Website Fingerprinting\\over TLS Handshake}

\author{\IEEEauthorblockN{Aida Ramezani, Amirhossein Khajehpour, Mahdi Jafari Siavoshani}
\IEEEauthorblockA{Information, Network, and Learning Lab\\Department of Computer Engineering\\
Sharif University of Technology\\
aidaramezani9776@gmail.com, amirhosseinkh@ce.sharif.edu, mjafari@sharif.edu}
}

%


\maketitle

\begin{abstract}
Analyzing users' Internet traffic data and activities has a certain impact on users' experiences in different ways, from maintaining the quality of service on the Internet and providing users with high-quality recommendation systems to anomaly detection and secure connection. Considering that the Internet is a complex network, we cannot disintegrate the packets for each activity. Therefore we have to have a model that can identify all the activities an Internet user does in a given period of time. In this paper, we propose a deep learning approach to generate a multi-label classifier that can predict the websites visited by a user in a certain period. This model works by extracting the server names appearing in chronological order in the TLSv1.2 and TLSv1.3 \textsl{Client Hello} packets. We compare the results on the test data with a simple fully-connected neural network developed for the same purpose to prove that using the time-sequential information improves the performance. For further evaluations, we test the model on a human-made dataset and a modified dataset to check the model's accuracy under different circumstances. Finally, our proposed model achieved an accuracy of 95\% on the test dataset and above 90\% on both the modified dataset and the human-made dataset.
\end{abstract}


%
\IEEEpeerreviewmaketitle

\section{Introduction}


Internet traffic classification is used for multiple intentions. Primarily, website classification was used for enterprise network management \cite{lght-class, comp-ser}, providing users with Quality-of-Service (QoS) \cite{comp-ser, 4446907, qos-tc, roughan2004class}, recommender systems based on users' activity \cite{6596414}, and anomaly detection \cite{kou2004survey}. These classification tasks can allow an adversary to use a website fingerprinting attack to invade the network's confidentiality. Recently, with the widespread usage of TLS encrypted protocol over the Internet, complex procedures are being employed to address the new challenges of encrypted traffic classification \cite{wang2017end, lotfollahi2020deep, mahdavi2018encrypted, zou2018encrypted}.

Traditionally, the port-based analysis was used to detect different applications over the network. However, with the emerging utilization of dynamic ports, such approaches have become obsolete. Other methods try to inspect the payload of packets (\ie, known as deep packet inspection) to identify the application or the website. These methods work perfectly for unencrypted packets. The most advanced approach is to use machine learning (ML) and deep learning models to extract the important features of the network traffic, either by experts or automatically, and to train an ML model to predict the application or the website used or visited.

For instance, Wang~\etal propose a classification method for encrypted traffic data that uses a 1-dimensional convolutional neural network \cite{wang2017end}. They use this model to automatically extract useful features from the raw ISCX traffic dataset and classify both VPN and nonVPN traffic data on 14 classes representing different activities, \eg, email, and streaming. Similar to this work, \cite{lotfollahi2020deep} introduces Deep Packet, a deep learning based approach that uses the same dataset as  \cite{wang2017end}, but with IP addresses masked, and use a convolutional neural network to recognize both traffic characterization and application identification. 

Alshammari~\etal~use a machine learning approach to detect encrypted from unencrypted tunnels. They use different learning algorithms trained on both packet header features, and statistical flow features, \eg, number of packets in forward/backward direction, without using the IP addresses, ports, and payloads on two encrypted traffic tunnels, SSH, and Skype to distinguish different services \cite{alshammari2011can}. This work proves that much can be inferred from the statistical features of the traffic flow. On statistical feature selection techniques, Shen~\etal~use the cumulative length of a sequence of packets as the traffic feature to fingerprint different webpages of a website with a K-NN classifier. They use encrypted traffic as raw data, but their approach is different from ours as their goal is to classify different webpages of one particular website \cite{shen2019webpage}. Here, we try to distinguish different websites from each other, having users visit different random webpages of each website. Salman~\etal~develop a multi-level classifier using ConvNet architecture, along with other deep learning classification methods to detect various applications on different network requirements of QoS and security policies \cite{salman2018multi}. They use a hierarchical approach to classification, starting from classes of interactive, streaming, bulk data transfer, and the transaction traffic to subcategories such as video calls, voice calls, and texting. The multi-level classifier can then continue to detect different applications and devices. They use a feature selection method and extract the size, interarrival time, direction, and transport protocol of the first packets of the flow.

In this paper, we use a feature selection approach and combine it with deep learning to develop a classifier that can predict the websites visited by a user. Our proposed website prediction aims to investigate whether the TLS handshake protocol guarantees users with enough confidentiality or an adversary or a recommender system can infer the user's activity from the TLS handshaking. 

\begin{figure*}[]
	\centering
	\includegraphics[width=\textwidth]{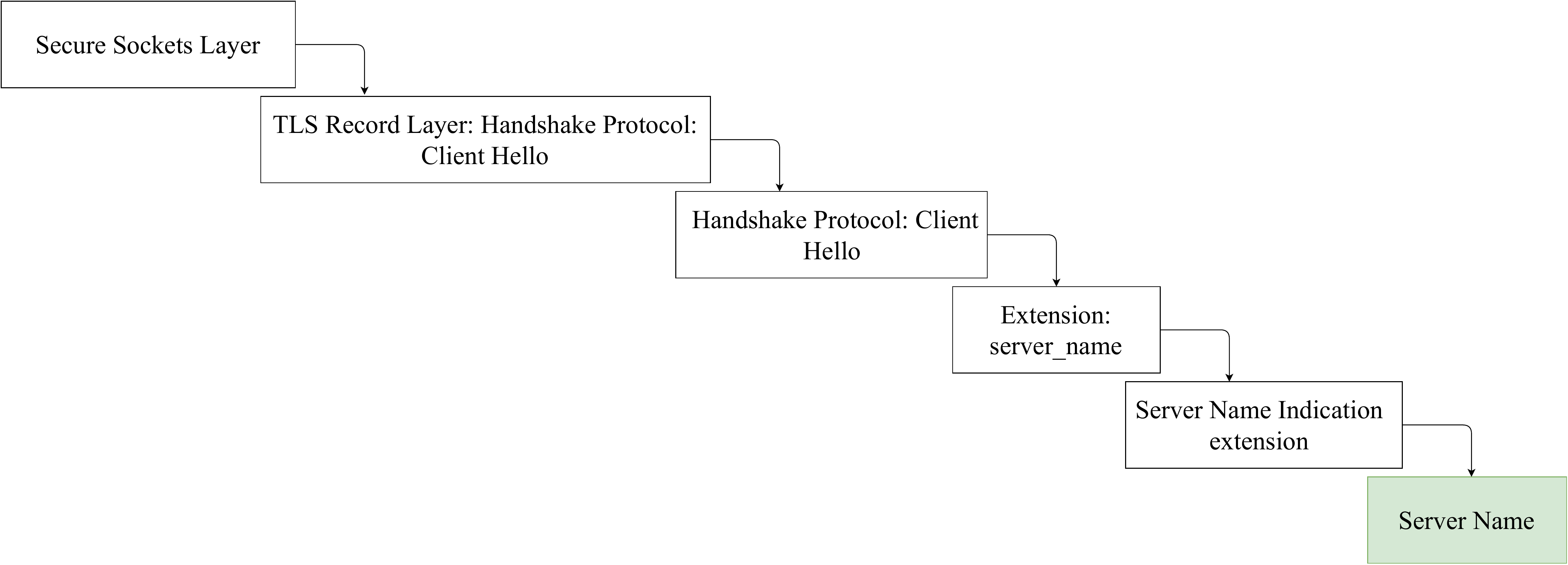}
	\caption{The hierarchical structure of \textsl{Client Hello} packet of TLS handshake protocol to reach the server name field.}
	\label{fig:packet}
\end{figure*}

In this regard, we extract the server name field as the main feature of packets. It is an unencrypted field in the \textsl{Client Hello} packet of the handshake protocol used in TLSv1.2 and TLSv1.3. 
Using this feature mitigates the need to analyze the encrypted traffic to discover any leakage of information. Figure~\ref{fig:packet} shows the structure of the \textsl{Client Hello} packets and the positions of the server name field within. We take a similar approach to \cite{zou2018encrypted}, and use the time sequential quality of the traffic flow for our classification task. They use a convolutional neural network combined with a Long Short-Term Memory (LSTM) model to extract the packet-level and flow-level information and use that information to distinguish the application class of the traffic flow.

\begin{figure*}[]
	\centering
	\includegraphics[width=\textwidth]{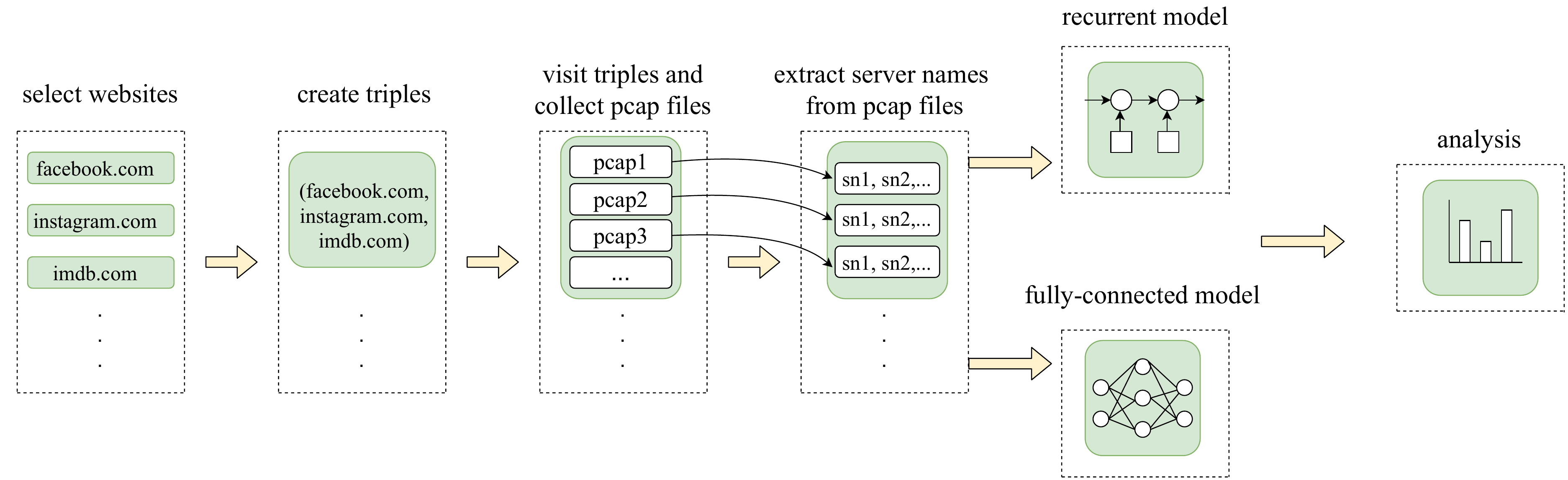}
	\caption{Our framework for website prediction task.}
	\label{fig:framework}
\end{figure*}

In this work, we try to identify the set of websites visited by a user from a given traffic flow using Long Short-Term Memory (LSTM) \cite{hochreiter1997long}. Our model works by using any 20 consecutive \textsl{Client Hello} packets from the flow and predicts the websites responsible for the generated traffic. By using the server name, this work can be considered as the follow-up work of \cite{ssl-inl}. The model provided in \cite{ssl-inl} predicts the server names from an encrypted TLS application data. As a result, by combining the model presented in this work with the technique presented in \cite{ssl-inl}, one can predict the user activity (\eg, visited websites) using only encrypted data of TLS flow, without any additional information from unencrypted parts of packet, proving that the newly found flaws by \cite{ssl-inl} on the TLS protocol is critical.
Figure~\ref{fig:framework} presents our framework for website prediction.

Despite the wide use and development of Internet traffic classification, to the best of our knowledge, there has been no method that can name all the activities generating a traffic flow. Considering the great possibility of having users multitask on the Internet, \eg, visiting multiple websites at a time, we must develop a model that considers this possibility and classifies a set of possible labels as the classification task. In this work, we address this issue by developing a multi-label classifier trained on a dataset labeled with more than one website.

The key contributions of this paper are summarized as follows:
\begin{itemize}
	\item Our proposed method does not need feature extractions, nor deals with the encrypted payload. The only feature used is the unencrypted server name field in the TLS protocol.
	\item We consider the sequential characteristics of the Internet traffic flow using a recurrent neural network to extract the time sequence information.
	\item In this paper, we tackle the problem of multitasking behavior of the users. We develop a multi-label classifier that predicts a set of websites visited at the same time. Applying our method for any initial choice of $n$ websites requires only $O(n^2)$ labels.
%
	\item Our experimental results are comparable to those of state-of-the-art works. Our proposed model has achieved an accuracy of $95.5$\% on the test dataset. 
	\item The dataset gathered for this work contains the newest versions of the TLS protocol applied in real-world settings.
\end{itemize}

The rest of the paper is organized as follows. Section \ref{section:data} introduces our data collection method and the data-set. Sections \ref{sec:method} and \ref{section:experiments} elaborate on the method architecture, the experiments, and their results. Finally, the conclusion sums up the paper's purpose and the method proposed.

 

\section{Data Collection}
\label{section:data}
A classifier's performance highly depends on the adequacy of the training data. To collect the dataset, we used a web-crawling process, and gathered the .pcap files generated from the traffic flow of visiting a set of websites. Here we explain our data collection approach and the datasets generated.

\subsection{Data collector system}
To create our dataset, we used 20 websites based on the Alexa list of most visited websites on the Internet. To generate the .pcap files, we used a \textit{Selenium}-based crawler to visit random entry pages of selected websites. The \textit{tcpdump} tool is used to generate the .pcap files created while visiting the pages. Each .pcap file was then reduced to the server name field of \textsl{Client Hello} packets of TLSv1.2, or TLSv1.3 handshake protocol appeared in it chronologically. As a result, for each initial .pcap file, we make a list of server names. Server names in \textsl{Client Hello} TLS handshake packets are unencrypted; thus, they are useful to extract information about the flow.

\subsection{Datasets}
In this section, we present our three datasets collected for the multi-tasking website prediction problem. The first dataset called the \emph{main dataset}, consists of $\binom{20}{2}$ labels, each label representing a possible set of three websites visited simultaneously. The second dataset, as we call it the \emph{supplementary dataset}, contains more than 800 labels chosen randomly from all possible  $\binom{20}{3}$ triples, to evaluate the model's performance on never-seen samples. The final dataset, or the so called \emph{real user dataset}, consists of three sub-datasets that are the traffic collected by real user activity on the Internet. We use this dataset to assess the model in a real-world environment circumstance. Here, we describe each dataset in details.

\subsubsection{Main Dataset}
For the main dataset used in train and test, we created $190$ labels, each consisting of three websites from the twenty selected websites. These 190 labels represent all $\binom{20}{2}$ ways that any two websites can be visited simultaneously by a user. More specifically, these 190 labels are chosen so that every two websites appear at least once in a triple. For each triple, the crawler visited 15 random entry pages of each website simultaneously and saved the .pcap file created. As a result, each .pcap file consists of all packets transferred through the network while the crawler was visiting multiple pages of three different websites at a time. This procedure is repeated 11 times for each triple. 

\subsubsection{Supplementary Test Dataset}
To test the model's performance on the triples not seen during the training process, we choose 843 triple labels randomly of all $\binom{20}{3}$ possible labels we could assign to a set of three websites. We follow the same procedure used for the main dataset and capture a .pcap file for each label.

\subsubsection{Real User Dataset}
In addition to the test data gathered with a crawler, to evaluate the model on a traffic flow collected by a real user, we assembled three sets of datasets by searching a set of websites on a computer and clicking on random public pages of the websites, simulating what a real user would do surfing multiple websites at a time. Then, we collected a .pcap file for each label.

For the first real-user-made test dataset, we visited one website at a time and gathered the .pcap file generated. For the second real-user-made test dataset, we visited two websites at a time and collected the .pcap files, and the final real-user-made test data was collected by visiting four different websites at a time to evaluate the model on samples that were labeled with more than three websites. For the rest of the paper, we will call these datasets uni-label, binary-label, and multi-label datasets, respectively. Table~\ref{tab:0} summarizes the information about each dataset.

\begin{table*}
\centering
\caption{Label types and the number of labels selected for each dataset.}
\begin{tabular}{ |c| c| c|}
 
 \hline
    \textbf{Dataset} & \textbf{Label Type} & \textbf{Number of Labels}\\
 \hline \hline
	Main dataset & Triple & 190 \\
	\hline
	Supplementary & Triple &  843\\
	\hline
	Uni-label real-user dataset & Single & 10\\
	\hline
	Binary-label real-user dataset & Tuple & 9\\
	\hline
	Multi-label real-user dataset & 4-Tuple & 7\\
	\hline
\end{tabular}

\label{tab:0}
\end{table*}

\section{Methodology}
\label{sec:method}
\subsection{Representation of the traffic flow}
To train our classifiers, we use the one-hot encoding to represent both server names and single websites. For $n$-tuple labels, \ie, tuple, triple, and 4-tuple, we use a vector of a size of 20, \ie, the number of selected websites, with the websites in the triple as ones, and the rest as zeros.


\subsection{Proposed models}
\begin{figure}[h]
	\centering
	\includegraphics[width=.85\linewidth]{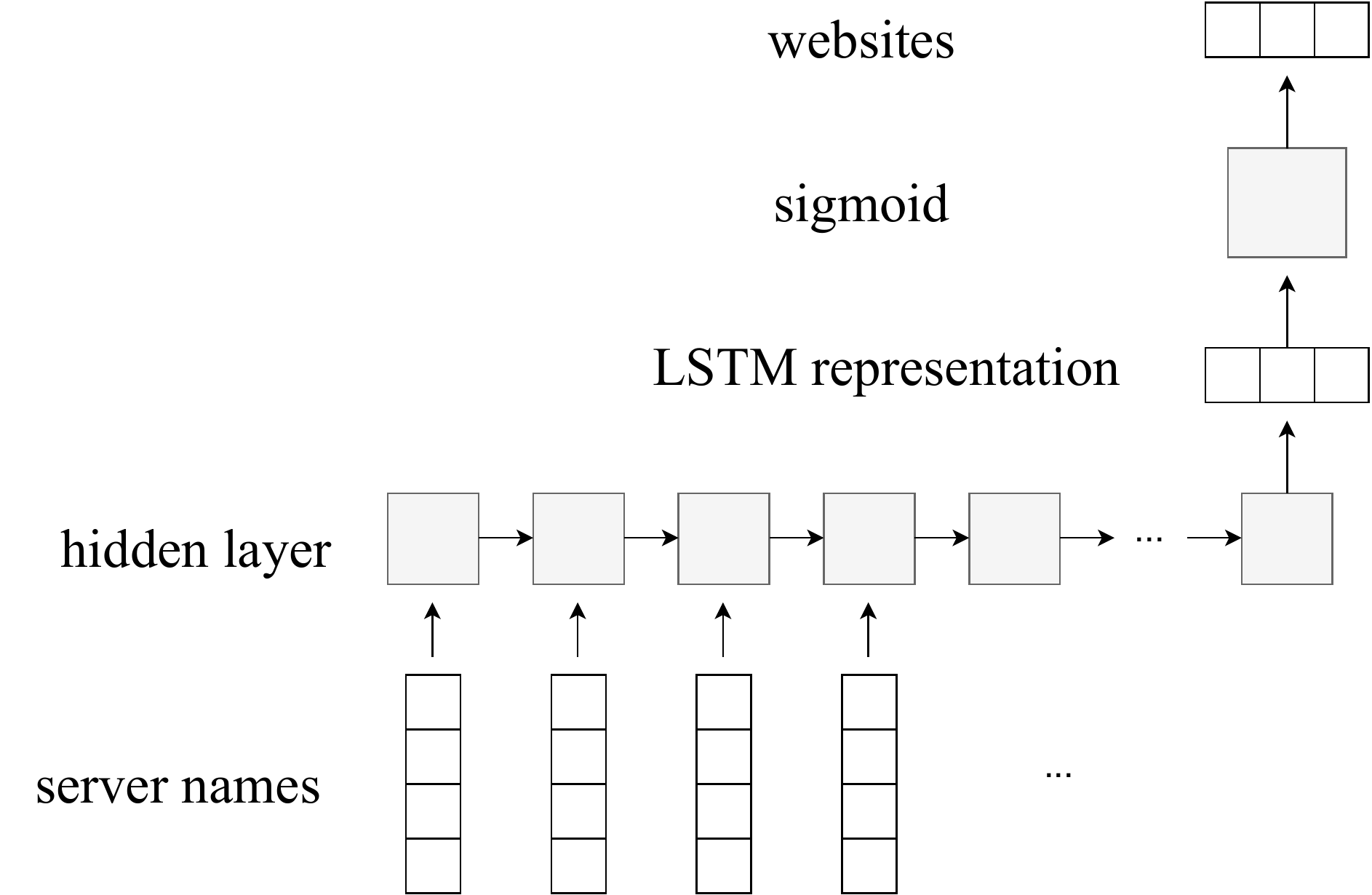}
	\caption{The general illustration of the proposed LSTM model.}
	\label{fig:model}
\end{figure}

We label each sequence of chronological server names with a label, \eg, a triple for the main dataset.  We devise two multi-label classifier models that predict a set of websites for a given sequence of server names and compare their performance over the test dataset.

\subsubsection{LSTM Model}
The first classifier is a recurrent neural network model consisting of a hidden layer of 256 LSTM cells. These types of recurrent neural networks have proven to be a powerful tool in working with sequential data. All the initial LSTM cell states are zero, and the time sequence used is  $T = 20$, and a learning rate of $0.01$. We take an early stopping approach to prevent the over-fitting of the model. To tackle the problem of simultaneous website visiting, the final layer of the model works as a multi-label classifier by using sigmoid cross-entropy cost function on the logits and Adam Optimizer \cite{kingma2014adam}. Figure~\ref{fig:model} gives a visual representation of this model.

\subsubsection{Fully-Connected Model}
To prove our point that the information provided by the packets' sequence is beneficial, we developed a regular fully-connected neural network as our alternative model. This model consists of 3 hidden layers with $256$, $218$, and $64$ neurons. A dropout rate of $0.2$ is used to avoid overfitting. Similar to the LSTM model, we use a sigmoid cross-entropy cost function to have a multi-label classifier. The training procedure runs for 50 epochs.




\section{Experiments}
\label{section:experiments}
%

We split the main dataset randomly into train and test sets with 85\% and 15\% portions of the dataset, respectively. We also use two additionals test datasets explained in Section~\ref{section:data} to evaluate our model. To train the LSTM model, each time we randomly choose a label (\ie, a .pcap file regarding that label) and a starting index in the set of consecutive server names appeared in the .pcap file. We then transform the sequence of server names to their one-hot representation and feed this sequence to the network. To measure the accuracy, we calculate the average number of correctly predicted, either existing or non-existing websites in the label. The accuracy is estimated as follows:

\begin{equation}
\label{eq:accuracy} 
\mathrm{Accuracy} = \frac{\mathrm{TP} + \mathrm{TN}}{\mathrm{TP} + \mathrm{FP} + \mathrm{TN} + \mathrm{FN}},
\end{equation}
where TP, FP, TN, and FN stand for True Positive, False Positive, True Negative, and False Negative respectively. 


To train the fully-connected model, each .pcap file is represented by the server names frequency vector that appeared in it. Using this representation, the traffic flow information, i.e., the server names' chronological order, is lost. All the layers, except for the last layer of this model, use the ReLU activation function. The accuracy of this model is estimated similarly to the LSTM model described in \eqref{eq:accuracy}

\begin{figure}[h]
	\includegraphics[scale=0.235]{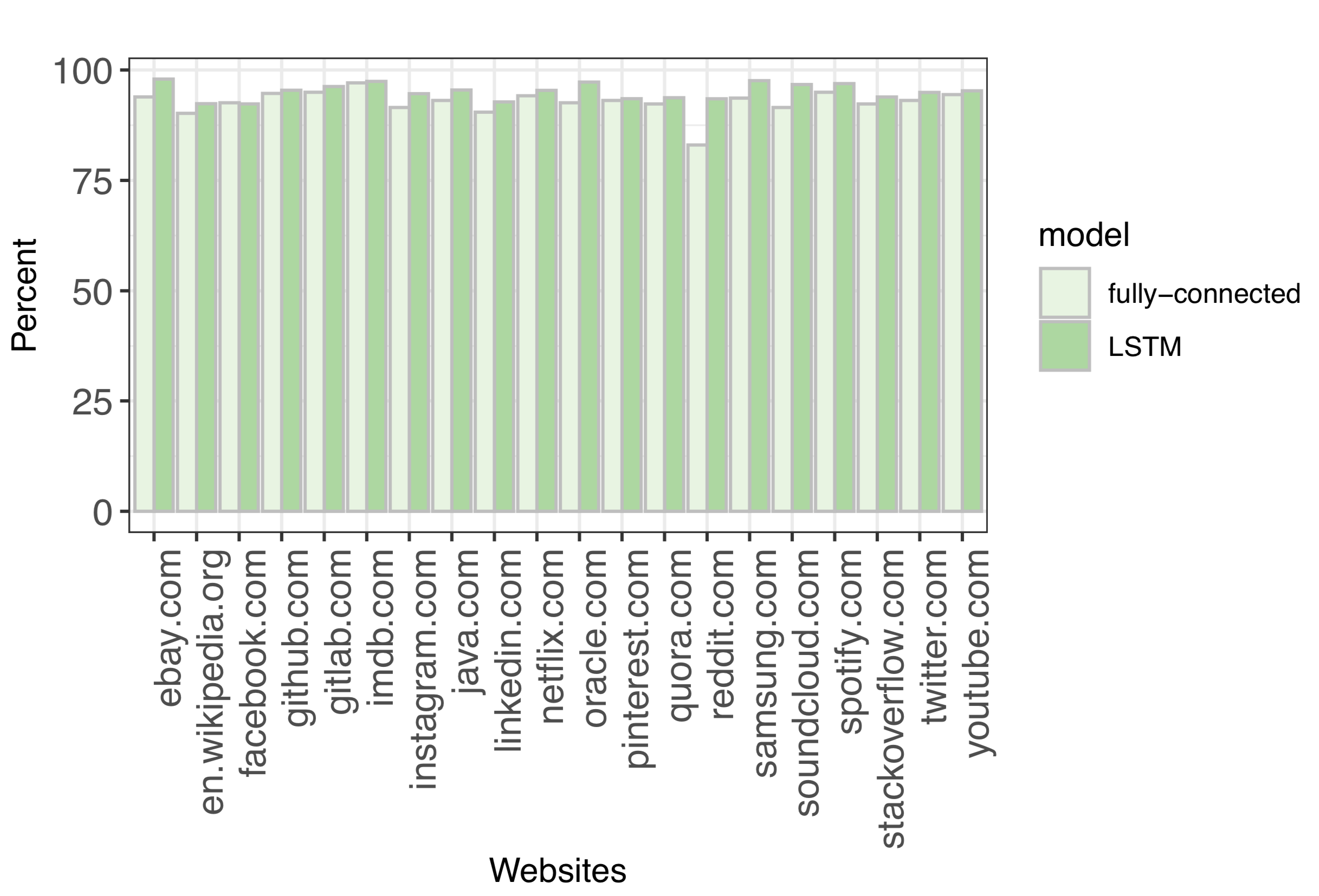}
	\caption{Comparing the achieved accuracies on the test dataset by the recurrent neural network model, and the fully-connected model.}
	\label{fig:accuracy of websites}
\end{figure}

\begin{figure}[h]
	\includegraphics[scale=0.6]{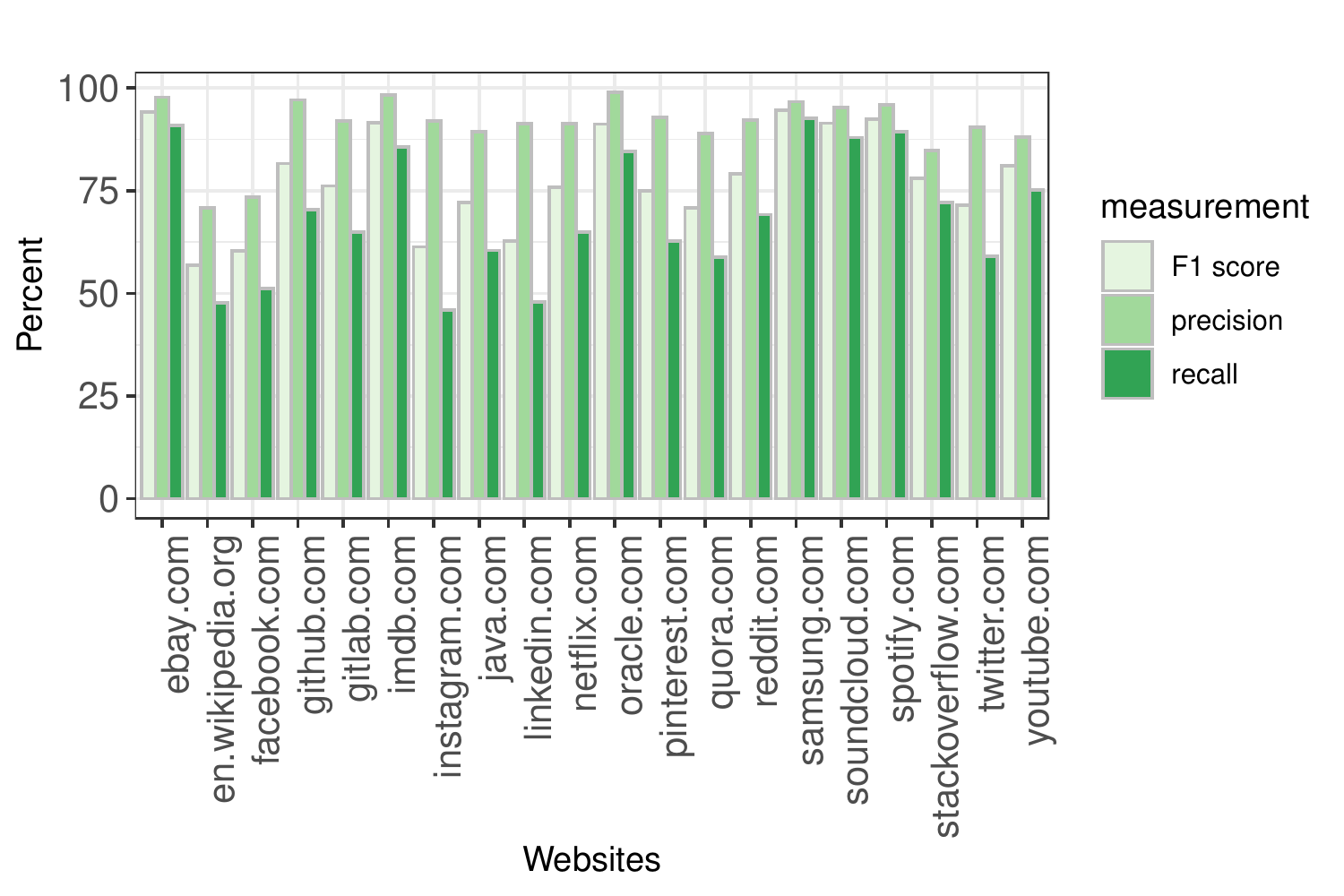}
	\caption{Precision, recall, and F1 score of the recurrent neural network model on the test dataset.}
	\label{fig:F1 of websites}
\end{figure}

Table~\ref{tab:acc} compares the performance of these two models on the main dataset test set. The LSTM model and the fully-connected model have achieved accuracies of 95.5\% and 92.7\%, respectively. Figure~\ref{fig:accuracy of websites} compares the accuracy of each website class on the main test set. The significant difference  between the achieved accuracies of the two models indicates that the server name appearing in a .pcap file have to be viewed from a time-sequence perspective, rather than eliminating the inherent time-sequence quality of the traffic flow ($ p =\, 0.0001804 \, \textless \, 0.01, \, paired \; t-test$). Figure~\ref{fig:F1 of websites} displays the recall, precision, and F1 score regarding each website of the LSTM model.

\begin{table*}[t]
\begin{center}
 \begin{tabular}{ |c || c c| c || c c |c |}
 \hline
   {\textbf{Websites}}& \multicolumn{3}{c||} {\textbf{LSTM model}} & \multicolumn{3}{c|} {\textbf{Fully-Connected Model}}\\ 
 \hline 
- & recall & precision & accuracy & recall & precision & accuracy \\
 \hline \hline
imdb.com	&0.856	&0.983	&\textbf{0.974}	&0.838	&0.861	 & 0.97  \\
\hline
github.com	&0.704	&0.971	 &\textbf{0.954}	&0.654	&0.944	 &0.947\\\hline
stackoverflow.com	&0.721	&0.848	&\textbf{0.939}	&0.706	&0.72  &0.923\\\hline
samsung.com	  &0.926	&0.967	&\textbf{0.976}	&0.864 	&0.679	  &0.937\\\hline
pinterest.com	&0.628	&0.929	&\textbf{0.935}&0.745	&0.774	  &0.931\\\hline
linkedin.com	&0.479	&0.914	&\textbf{0.928}	 &0.469	&0.441  	&0.905\\\hline
soundcloud.com	 &0.88	&0.953	&\textbf{0.967}	 &0.694	&0.768  	&0.915\\\hline
instagram.com	&0.46	&0.92	 &\textbf{0.946} &0.483	&0.933	&0.915\\\hline
java.com	&0.605	 &0.893	&\textbf{0.955}	&0.673	&0.767	 &0.931\\\hline
gitlab.com	 &0.65	&0.919	&\textbf{0.963}	&0.694	&0.895	 &0.95\\\hline
quora.com	&0.588	&0.889	&\textbf{0.937}	&0.771	&0.673	  &0.923\\\hline
spotify.com	 &0.893	&0.959	&\textbf{0.969	}&0.887	&0.783	  &0.95\\\hline
oracle.com 	&0.845	&0.99	&\textbf{0.973}	&0.681 	&0.711	  &0.926\\\hline
ebay.com 	&0.91	&0.978	&\textbf{0.979}	&0.778	&0.729 &	0.939\\\hline
en.wikipedia.org	&0.476	&0.707	&\textbf{0.924}	&0.391	&0.667  	&0.902\\\hline
reddit.com	  &0.691	&0.922	&\textbf{0.935	}&0.791 &0.515	  &0.83\\\hline
twitter.com  	&0.59	&0.905	&\textbf{0.949}	&0.641	&0.676   	&0.93103\\\hline
youtube.com	  &0.752	&0.881	&\textbf{0.953	}&0.673	&0.868  	&0.944\\\hline
facebook.com	&0.511	&0.735	&0.923	&0.6   &	0.846	 & \textbf{ 0.926}\\\hline
netflix.com 	&0.649	&0.913	&\textbf{0.954}	&0.704	 &0.864	& 0.942\\\hline

\end{tabular}
\caption{Classification recalls, precisions, and accuracies for the main test dataset of the LSTM model, and the fully-connected model. }
\label{tab:acc}
\end{center}

\end{table*}

%
%
%

\begin{table*}
\begin{center}
 \begin{tabular}{ |c | c | c|}
 \hline
   \textbf{Dataset} & \textbf{LSTM model} & \textbf{Fully-Connected Model}\\
 \hline 
	Uni-label real-user dataset & 0.919 & 0.6 \\\hline
	Binary-label real-user dataset & 0.948 & 0.864\\
	\hline 
	Multi-label real-user dataset & 0.872 & - \\ \hline
\end{tabular}

\caption{The accuracies on the real-user dataset of the LSTM, and the fully connected models.}
\label{tab:2}
\end{center}

\end{table*}


%

To measure the performance of the models given a real user, we tested it on the real-made datasets addressed in section~\ref{section:data}.  Table~\ref{tab:2} compares the achieved accuracies of the LSTM and the fully-connected models on the uni-label real-user and the binary-label real-user datasets. The obtained results prove that the LSTM model can successfully predict a set of websites a real user is visiting simultaneously in a complex network, and be used in real-world environments. We tested only the LSTM model on the real-user-multi-label dataset, as it had out-performed the fully-connected model on all previous datasets. The LSTM model achieved an accuracy of 87.18\% on the multi-label dataset. On average, it can predict 2.9 out of 4 labels correctly.

To further assess the LSTM model's performance on a set of new labels that are not seen during the training process, we employ the supplementary dataset collected in Section~\ref{section:data}. The LSTM model could successfully achieve an accuracy of 93.21\% on this dataset, which contained new labels. This result can be explained by knowing that the main dataset labels are selected so that all 190 possible ways of having any two websites in a triple were covered.

For final performance measurement, we remove server names that included the websites' names in them from the main test set. For example, server name $1sn34.ebay.com$ was removed from its respective .pcap files since it contained $ebay.com$. The dataset still kept its sequential quality, but some server names were removed from the flow. In this scenario, the LSTM model achieved an accuracy of 93.82\%. This result emphasizes that the model can predict the websites visited by the user, even when the server names do not leak any specific information about the websites' names.

%
%
%

\section{Conclusion}
In this paper, we proposed a novel framework that classifies internet packets to discover users' activities surfing on the internet. To the best of our knowledge, our framework is the first to predict more than one activity done by the users at a time, considering the multitasking inherent nature of working with the internet. We use an unencrypted feature of the TLS handshake protocol, and a recurrent neural network model with LSTM cells to extract the features of the packet flow in a time sequence. Experiment results have shown that considering the information from the sequential character of the traffic flow by using an LSTM model, outperforms looking at the whole flow at once. Furthermore, our framework can be used for any combination of websites since it needs no more than $O(n^2)$ triple labels. We conclude that the unencrypted information in the TLS handshake protocol packets used along with the sequential nature of the traffic flow can leak specific information about the users' activity that may be favored confidential.


\section*{Acknowledgment}
The authors would like to thank Saeed Aghamiri for his helpful discussions and feedback.

%

\bibliography{sections/bibliography.bib}
\bibliographystyle{ieeetr}

\end{document}